# Uncertainty Relation for the Radial Momentum and Radial Coordinate in the Coulomb Potential

A. M. Ishkhanyan[a,b] and V. P. Krainov[c,*]

[a] Russian-Armenian University, Yerevan, 0051 Armenia
[b] Tomsk Polytechnic University, Tomsk, 634050 Russia
[c] Moscow Institute of Physics and Technology (State University),
Dolgoprudnyi, Moscow oblast, 141700 Russia

**Abstract**—We have obtained the uncertainty relations for arbitrary states of the hydrogen atom. It is shown that the minimal value of the uncertainty relation ($\Delta p_r \Delta r \to \hbar/2$) is attained for the circular Rydberg states.

The uncertainty relations $\Delta p_x \Delta x \geq \hbar/2$ in 1D problems of quantum mechanics are well known [1]. This work aims at the derivation of analogous relations in the central potential and analysis of the most interesting case of the Coulomb potential. We denote by $R$ the normalized real-valued radial wavefunction of the bound state with quantum numbers $n$ and $l$. In the central potential, the Hermitian operator of the radial momentum has the form [2]

$$\hat{p}_r = -i\left(\frac{d}{dr} + \frac{1}{r}\right).$$

The Schrödinger equation for the radial function has the form ($\hbar = m = 1$ everywhere)

$$\hat{p}_r^2 R(r) = k^2(r)R(r), \quad k^2 = 2[E - U_{\text{eff}}(r)],$$
$$U_{\text{eff}}(r) = U(r) + \frac{l(l+1)}{2r^2}, \quad \int_0^\infty r^2 dr R^2(r) = 1. \quad (1)$$

Let us define the average value of the coordinate and its fluctuations:

$$\langle r \rangle = \int_0^\infty R^2(r) r^3 dr, \quad \delta r = r - \langle r \rangle. \quad (2)$$

We will prove that the mean value of the Hermitian operator of the radial momentum is zero. We will mark by prime the differentiation with respect to the radial coordinate. This gives

$$i\langle \hat{p}_r \rangle = \int_0^\infty r^2 dr R\left(R' + \frac{R}{r}\right) = \int_0^\infty r^2 dr R' R$$
$$+ \int_0^\infty r dr R^2 = -\int_0^\infty dr R(r^2 R)' + \int_0^\infty r dr R^2 \quad (3)$$
$$= -\int_0^\infty r^2 dr R\left(R' + \frac{R}{r}\right) = 0,$$

since the mean value of the radial momentum turned out to be equal to itself with the opposite sign. Therefore, the operator of the radial momentum fluctuation is equal to the momentum operator itself, $\delta \hat{p}_r = \hat{p}_r$.

Following the Weyl method [1], we consider the auxiliary integral

$$I(\alpha) = \int_0^\infty r^2 dr |(\alpha \delta r - i\hat{p}_r)R|^2 \geq 0. \quad (4)$$

It can be written in the real form:

$$I(\alpha) = \int_0^\infty r^2 dr \left\{\alpha R \delta r - R' - \frac{R}{r}\right\}$$
$$\times \left\{\alpha R \delta r - R' - \frac{R}{r}\right\} \geq 0. \quad (5)$$

We evaluate three individual terms of this integral:

$$I_1(A) = \alpha^2 \int_0^\infty r^2 dr R^2(r) (\delta r)^2 = \alpha^2 \langle \delta r^2 \rangle. \quad (6)$$

Further, we evaluate

$$I_2(\alpha) = \int_0^\infty r^2 dr \left(R' + \frac{R}{r}\right)^2 = \int_0^\infty dr\{r^2 R'^2 + 2rRR' + R^2\}$$

$$= \int_0^\infty dr\{-R(r^2 R')' + 2rRR' + R^2\} \quad (7)$$

$$= \int_0^\infty dr\{-r^2 RR'' + R^2\} = -\int_0^\infty r^2 dr R\left(R'' + \frac{2}{r}R'\right) = \langle \delta \hat{p}_r^2 \rangle.$$

In our calculations, we assumed that

$$\int_0^\infty rdr RR' = -\int_0^\infty dr R(rR)' = -\int_0^\infty rdr RR' - \int_0^\infty dr R^2,$$
$$2\int_0^\infty rdr RR' = -\int_0^\infty dr R^2. \quad (8)$$

Finally, the third integral has the form (with allowance for expression (3))

$$I_3(\alpha) = -2\alpha \int_0^\infty r^2 dr \left\{R(r - \langle r \rangle)\left(R' + \frac{R}{r}\right)\right\}$$
$$= -2\alpha \int_0^\infty r^3 dr R\left(R' + \frac{R}{r}\right) = -2\alpha \int_0^\infty r^3 dr RR' - 2\alpha. \quad (9)$$

Since

$$\int_0^\infty r^3 dr RR' = -\int_0^\infty dr R(r^3 R') = -\int_0^\infty r^3 dr RR' - 3,$$
$$\int_0^\infty r^3 dr RR' = -\frac{3}{2}, \quad (10)$$

we find from expression (9) that $I_3(\alpha) = \alpha$.

Therefore, integral (5) is given by

$$I(\alpha) = \alpha^2 \langle \delta r^2 \rangle + \alpha + \langle \delta \hat{p}_r^2 \rangle \geq 0. \quad (11)$$

This expression can be written in the form

$$\left\{\alpha\sqrt{\langle \delta r^2 \rangle} + \frac{1}{2\sqrt{\langle \delta r^2 \rangle}}\right\}^2 - \frac{1}{4\langle \delta r^2 \rangle} + \langle \delta \hat{p}_r^2 \rangle \geq 0, \quad (12)$$

which gives the uncertainty relation

$$\langle \delta \hat{p}_r^2 \rangle \langle \delta r^2 \rangle \geq \frac{\hbar^2}{4}. \quad (13)$$

Let us define the wavefunction for which expression (13) becomes an equality. Using inequality (12), we obtain

$$\alpha = -\frac{1}{2\langle \delta r^2 \rangle}. \quad (14)$$

In accordance with relations (4) and (14), we obtain the following equation:

$$\left(\frac{r - \langle r \rangle}{2\langle \delta r^2 \rangle} + \frac{d}{dr} + \frac{1}{r}\right) R(r) = 0. \quad (15)$$

Its solution has the form

$$R(r) = \frac{C}{r} \exp\left[-\frac{(r - \langle r \rangle)^2}{4\langle \delta r^2 \rangle}\right]. \quad (16)$$

It has a singularity at the origin. This singularity can be ignored in principle if $\langle r^2 \rangle \gg \langle \delta r^2 \rangle$ so that the exponential smallness compensates the power-law increase.

Let us now consider the state of the hydrogen atom. For an arbitrary state $nl$ of the hydrogen atom, the variance of the coordinate is given by

$$\langle \delta r^2 \rangle = \frac{n^2(n^2 + 2) - l^2(l + 1)^2}{4}. \quad (17)$$

The radial matrix elements of the coordinate and its square can be calculated by the method of the generalized Kramers virial theorem [3].

Let us now consider the variance of the radial momentum. From relation (1), we obtain

$$\langle \delta \hat{p}_r^2 \rangle = \langle \hat{p}_r^2 \rangle = 2\langle E_n \rangle + 2\left\langle \frac{1}{r} \right\rangle - l(l + 1)\left\langle \frac{1}{r^2} \right\rangle$$
$$= \frac{1}{n^2} - \frac{2l(l + 1)}{n^3(2l + 1)}.$$

Here, the matrix elements of the inverse powers of the coordinate can be calculated elementary [4]. For the product of indeterminacies, we obtain

$$\langle \delta r^2 \rangle \langle \delta \hat{p}_r^2 \rangle = \left\{\frac{n^2 + 2}{4} - \frac{l^2(l + 1)^2}{4n^2}\right\}\left[1 - \frac{2l(l + 1)}{n(2l + 1)}\right]. \quad (18)$$

This expression is minimal for the maximal value of orbital quantum number $l = n - 1$ (circular orbits):

$$\langle \delta r^2 \rangle \langle \delta \hat{p}_r^2 \rangle_{\min} = \frac{2n + 1}{4(2n - 1)} > \frac{1}{4}. \quad (19)$$

For $n \to \infty$, it indeed tends to the limit of 1/4. As noted above, this statement follows from the inequality

$$\langle r \rangle^2 = n^4 \gg \langle \delta r^2 \rangle = n^3/2, \quad n \gg 1. \quad (20)$$

Thus, the minimal value $\hbar^2/4$ in the uncertainty relation for the classical states is attained for $\langle r^2 \rangle \gg \langle \delta r^2 \rangle$.

ACKNOWLEDGMENTS

This study was supported by the Armenian State Scientific Committee (grant no. 18RF-139), Armenian National Foundation for Science and Education (grant no. PS-4986), the Russian–Armenian (Slavonic) University, the Ministry of Education and Science of the Russian Federation (project no. 3.873.2017/4.6), and the Russian Foundation for Basic Research (project no. 18-52-05006).

REFERENCES

1. L. D. Landau and E. M. Lifshitz, *Course of Theoretical Physics,* Vol. 3: *Quantum Mechanics: Non-Relativistic Theory* (Fizmatlit, Moscow, 2016; Pergamon, New York, 1977).
2. S. Flugge, Practical Quantum Mechanics (Springer, Berlin, Heidelberg, 1999; Mir, Moscow, 1974), Vol. 1, Probl. No. 59.
3. V. V. Kiselev, *Quantum Mechanics* (Mosk. Tsentr Nepreryv. Mat. Obraz., Moscow, 2009) [in Russian].
4. Yu. M. Belousov, S. N. Burmistrov, and A. I. Ternov, *Problems in Theoretical Physcis* (Intellekt, Dolgoprudnyi, 2013) [in Russian].